\documentclass{article}
\usepackage{fullpage,times,amsfonts}

\begin{document}

\newtheorem{thm}{Theorem}
\newtheorem{prop}[thm]{Proposition}
\newtheorem{cor}[thm]{Corollary}
\newtheorem{lemma}[thm]{Lemma}
\newtheorem{defn}[thm]{Definition}
\newtheorem{claim}[thm]{Claim}

\newcommand{\K}{{\cal K}}
\newcommand{\Tr}{{\rm Tr\,}}
\newcommand{\A}{{\cal A}}
\newcommand{\acc}{{\rm ACC}}
\newcommand{\rej}{{\rm REJ}}
\newcommand{\ket}[1]{|{#1}\rangle}
\newcommand{\bra}[1]{\langle{#1}|}
\newcommand{\ketbra}[1]{\ket{#1}\,\bra{#1}}
\newcommand{\braket}[2]{\langle{#1}|{#2}\rangle}
\newcommand{\Hil}{{\cal H}}
\newcommand{\D}{{\cal D}}
\newcommand{\proof}{\noindent {\bf Proof:}}
\newcommand{\qed}{\hfill $\Box$}
\newcommand{\C}{{\cal C}}
\newcommand{\E}{{\cal E}}
\newcommand{\F}{{\cal F}}
\newcommand{\Sc}{{\cal S}}

\title{Uncloneable Encryption}

\author{Daniel Gottesman\\Perimeter Institute\\35 King Street N\\Waterloo, Ontario N2J
2W9 Canada\\E-mail: dgottesman@perimeterinstitute.ca}
\date{}

\maketitle

\begin{abstract}
Quantum states cannot be cloned.  I show how to extend this
property to classical messages encoded using quantum states, a
task I call ``uncloneable encryption.''  An uncloneable encryption
scheme has the property that an eavesdropper Eve not only cannot
read the encrypted message, but she cannot copy it down for later
decoding.  She could steal it, but then the receiver Bob would not
receive the message, and would thus be alerted that something was
amiss.  I prove that any authentication scheme for quantum states
acts as a secure uncloneable encryption scheme.  Uncloneable
encryption is also closely related to quantum key distribution
(QKD), demonstrating a close connection between cryptographic
tasks for quantum states and for classical messages. Thus,
studying uncloneable encryption and quantum authentication allows
for some modest improvements in QKD protocols. While the main
results apply to a one-time key with unconditional security, I
also show uncloneable encryption remains secure with a
pseudorandom key.  In this case, to defeat the scheme, Eve must
break the computational assumption behind the pseudorandom
sequence before Bob receives the message, or her opportunity is
lost.  This means uncloneable encryption can be used in a
non-interactive setting, where QKD is not available, allowing
Alice and Bob to convert a temporary computational assumption into
a permanently secure message.
\end{abstract}

\section{Introduction}

The primary application of cryptography is to send secret messages.
Typically, there is a sender Alice who wishes to communicate with a
receiver Bob, but an eavesdropper Eve is listening to their messages.
To stop her, Alice will encrypt anything she wishes to say.  In order
to decrypt these messages, Bob must possess a secret key which is
unknown to Eve, giving him an advantage over Eve.  While Bob can
easily read the secret message, Eve, lacking the key, will find it
much harder to do so, or even impossible.  Another common task is
authentication of a message: Alice and Bob do not care if Eve reads
the message, but want to make sure she does not change it.  Naturally,
a message can be both encrypted and authenticated.

Many cryptographic systems are based on computational assumptions.
In this case, Eve's task of learning the contents of a message (or
even learning a single bit of information about the message) is
equivalent to solving some computationally difficult problem, one that
cannot be answered in polynomial time in some security parameter $s$.
These cryptosystems come in two flavors: symmetric systems, in which
Alice's encryption key is the same as Bob's decryption key; and
asymmetric systems, in which they are different.  Asymmetric systems
are usually public key systems in which Alice's encryption key is also
known to Eve.  However, Bob's key must always be secret, or else Eve
can decode the message just as easily as Bob.

Stronger security can be provided by the one-time pad.  In this
scheme, Alice and Bob share a secret key that is as long as the
message.  The ciphertext for the encrypted message consists of the
bits of the original plaintext {\sc XOR}ed with the corresponding bits
of the key.  When the key is only used once, this scheme is
unconditionally secure: lacking information about the key, Eve cannot
learn anything at all about the message, no matter what computational
power she has.

However, the following ``cloning'' attack will break even the one-time pad:
\begin{enumerate}
\item Alice and Bob share a secret (classical) key $k$.
\item Alice encrypts a (classical) message $m$ to Bob using an
encryption scheme with the key $k$.
\item Eve receives the message, and performs an attack of her choice.
In particular, she attempts to copy the encrypted message.
\item Eve passes the original message on to Bob, who checks if it has
been tampered with.
\item Eve acquires a copy of the secret key $k$.
\end{enumerate}
When the ciphertext is classical, Eve can copy the message without
changing it at all.  Therefore, when Bob receives the message, he does
not know Eve has interfered with the transmission.  Still, when Eve
learns the key in the final step, she can read the message without
difficulty.

For the one-time pad, perhaps this attack is not a severe worry.
Since the key is not needed again once the message is read, it
makes sense to destroy it immediately.  However, this is not as
easy as it sounds: typically the key must be stored for a while,
perhaps on a magnetic medium such as a hard drive.  Simply asking
a computer to ``delete'' a file only eliminates the directory
entry for the file. Actually erasing a file requires a separate
program to overwrite the bits of the file.  Even then, traces
remain which are accessible to forensic techniques.
Schneier~\cite{schneier}, for instance, recommends multiple
overwriting or physical destruction of magnetic media containing
expired keys.  The message, in contrast, can be deleted
immediately upon being read, and need never leave RAM.  It is thus
much easier to erase without leaving any traces behind.

The cloning attack is a more serious problem for computationally
secure schemes.  In such schemes, the same key is typically used
repeatedly, giving Eve more opportunity to steal it, enabling her to
read not only future messages, but past ones as well.  In fact, Eve
may not even need to steal the key: she can simply copy down the
message and begin a time-consuming brute-force attack on the
computational assumption.  Alternately, she could wait in the hopes
that future improvements in computer hardware or algorithms make it
easier to break the encryption scheme.  Furthermore, having the
ciphertext of many encrypted messages to examine is very helpful when
performing cryptanalysis.

Quantum states have the property that they cannot be copied
\cite{WZ,Dieks}. (See the textbook of Nielsen and Chuang~\cite{NC}
for an introduction to quantum information.) However, on the
surface, this only applies to superpositions of states. For
instance, if a classical message is sent using a standard set of
basis states, it can easily be copied without being disturbed. In
general, an unencrypted classical message can always be copied:
reading it constitutes making a copy.  However, when we move to
{\em encrypted} classical messages, the picture is very different.
By encrypting a classical message using quantum states, we can
produce {\em uncloneable encryption} schemes, which are secure
even against the cloning attack.  In particular, for any attack by
Eve, either she gets caught by Bob with high probability (for
instance, if she steals the message), or she gets almost no
information about the message.

Quantum mechanics has other cryptographic applications as well
(see~\cite{PT} for a survey).  The best-known is quantum key
distribution (QKD)~\cite{BB84}, which enables Alice and Bob to
create a secure classical secret key despite the potential
presence of an eavesdropper.  QKD requires only an insecure
quantum channel and authenticated (but unencrypted) classical
channels, but unfortunately requires multiple rounds of
back-and-forth communication between Alice and Bob.  The primary
proposed application of QKD is to create a secret key which is
then used with the one-time pad to send unconditionally secure
messages.  In contrast, uncloneable encryption is a noninteractive
protocol which can be used to enhance the security of the one-time
pad or computationally secure encryption schemes. Alternatively,
it can be used to perform QKD in a way requiring relatively little
interaction, and with other small improvements in efficiency and
security.  Noninteractivity is useful in a variety of contexts. It
is essential for storage of information, and important when there
is a substantial transmission lag (for instance, when
communicating with a space probe in the outer solar system).  It
is also very convenient for more mundane communication contexts
where there is a modest, but not completely negligible, time cost
for transmissions.  Uncloneable encryption is closely related to
some forms of QKD, but it can be best viewed as a stronger version
of symmetric encryption that shares the intrusion-detection
ability of quantum key distribution.  Another related concept is
that of ``secure direct communication,''~\cite{sdc} which is,
however, also explicitly interactive.  Curiously, the protocol
most closely related to uncloneable encryption is actually an
unpublished precursor~\cite{BBB} to BB84 (which I only learned of
after completing the first version of this paper), in which
messages are encrypted in a way similar to that described below.
The notion of uncloneability did not exist; instead, the purpose
of that protocol was to reuse the secret key if no intrusion was
detected.

Another side of the subject of quantum cryptography is to create
cryptographic protocols to protect quantum states.  For instance,
protocols for encryption~\cite{qenc1,qenc2} and
authentication~\cite{qas} of quantum states exist.  The quantum
authentication schemes, in particular, differ substantially from
classical schemes in that any quantum authentication scheme must
also encrypt a message.  It turns out that quantum authentication
protocols have precisely the properties needed to create
uncloneable encryption schemes.  In the past, cryptographic
protocols for quantum and classical messages have been viewed
largely as parallel subjects which have a good deal of similarity,
but little direct connection; this result establishes a more
intimate relationship, where a protocol for quantum states can be
used to perform a task protecting classical data.

In section~\ref{sec:def}, I give a technical definition of uncloneable
encryption, and show that any authentication scheme for quantum states
can be used as a secure uncloneable encryption scheme.  In
section~\ref{sec:QKD}, I go on to discuss the relationship between
uncloneable encryption and QKD.  Section~\ref{sec:prepandmeas} describes a
``prepare-and-measure'' uncloneable encryption scheme similar to the
well-known BB84 QKD protocol.  Such a scheme does not require
entangling quantum operations or a quantum memory, and therefore might
be experimentally implementable in the near future.  In
section~\ref{sec:compsec}, I show that uncloneable encryption remains
secure when the key is not a single-use item, but is instead generated
with a pseudorandom number generator based on a computational
assumption.  Finally, I conclude in section~\ref{sec:open} with some
remaining open questions.

\section{Uncloneable Encryption and Quantum Authentication}
\label{sec:def}

I will suppose throughout most of this paper that Alice and Bob share
a secret classical key $k \in \K$ which they will only use to send one
message.  If Alice wants to send a classical message $m$ to Bob, she
will use some encoding that depends on $k$; in general this could be a
mixed state $\sigma_k (m)$.  In order for this to be a good encryption
scheme, the transmitted density matrix, averaged over possible values
of the key, should not depend on the message:

\begin{defn}
\label{def:enc}
  Let $\sigma(m) = (\sum_k \sigma_k(m))/|\K|$.  Then $\sigma_k(m)$ is
  an (unconditionally secure) encryption scheme with error $\epsilon$
  if the trace distance $D(\sigma(m),\sigma(m'))=\frac 1 2 \Tr
  |\sigma(m)-\sigma(m')| \leq \epsilon$ for $m \neq m'$.
\end{defn}

That is, someone who does not know the key has essentially no
information about the message.  Definition~\ref{def:enc} only
addresses the secrecy of the message; for a useful protocol, we also
require that someone who does know the key is able to read the
message.  In fact, we usually restrict attention to encoding schemes
for which there are computationally efficient procedures to encode
$(m,k) \mapsto \sigma_k(m)$ and decode $(\sigma_k(m),k) \mapsto m$.

In order to have an uncloneable encryption scheme, we need an
additional condition.  A general attack by Eve is a superoperator $\A$
acting on $\sigma_k(m)$.  This represents the action Eve performs when
she first gets the encrypted state, before she learns the key, so $\A$
cannot depend on $k$.  The output of $\A$ can be divided into two
parts, a density matrix $\sigma_{{\rm Bob},k}(m)$ which is sent on to
Bob and the remainder which is kept by Eve.

To take the next step, we must assume Bob has some efficient checking
procedure $(\sigma_{{\rm Bob},k}(m), k) \mapsto \{\acc,\rej\}$ which
allows him to detect Eve's tampering.  He accepts the message when he
gets outcome $\acc$; when he gets outcome $\rej$ he concludes Eve may
have stolen the encrypted message, and he and Alice can take whatever
steps are necessary to protect themselves.  They may need to protect
the key $k$ especially well, for instance, or act to neutralize any
damage caused if Eve learns $m$.  We let $\rho_k (m)$ be Eve's
residual density matrix conditioned on the case that Bob gets outcome
$\acc$, and let $P_k(m)$ be the probability that Bob accepts the
message $m$.  In general, $\rho_k(m)$ and $P_k(m)$ can depend on the
attack $\A$.  For notational simplicity, I will hereafter write
$P_k(m)$ simply as $P(m)$.

\begin{defn}
\label{def:unclone}
An encryption scheme $\sigma_k(m)$ with error $\epsilon$ is an
uncloneable encryption scheme with error $\epsilon$ if, for any two
messages $m \neq m'$ and all attacks $\A$ by Eve, for a fraction of at
least $1-\epsilon$ of the possible values of the key $k$, the trace
distance $D\big(P(m)\rho_k(m),P(m')\rho_k(m')\big) \leq \epsilon$.
\end{defn}

Note that it is easy from this definition to prove two useful
properties: that $|P(m) - P(m')|$ is small and that, except when
$P(m)$ is very small, $D(\rho_k(m), \rho_k(m'))$ is small.  That is,
Eve's chance of being caught does not depend much on the message being
sent, and, unless she has a large chance of being caught, she has
little information about the message, even after learning the key.  In
particular, Eve cannot tell whether the message was $m$ or $m'$.  Note
that an uncloneable encryption scheme by definition must also encrypt
the message, so that Eve, even if she gets caught, cannot read the
message until she learns the key.  It is unclear whether this is an
independent condition, or whether it would follow from the
uncloneability requirement alone.  (A classical message sent
completely unencrypted can always be copied, but it may be possible to
create partially encrypted messages which are uncloneable.)

To create uncloneable encryption schemes, we can use quantum
authentication schemes.  A quantum authentication scheme is an
encoding that works on unknown quantum states: $(\ket{\psi},k) \mapsto
\sigma_k (\ket{\psi})$, where $\ket{\psi}$ is from some Hilbert space
$\Hil$.  This map should be a quantum operation (for instance, it must
be linear on the Hilbert space for $\ket{\psi}$) and should be
implementable efficiently.  Eve then performs an attack $\A$,
producing a state $\sigma_{{\rm Bob},k}(\ket{\psi})$ which is sent on
to Bob.  Bob then has an efficient decoding quantum operation
$\D(\sigma_{{\rm Bob},k}(\ket{\psi}), k)$.  The image of $\D$ is $\Hil_2
\otimes \Hil$, where $\Hil_2$ is a two-dimensional Hilbert space with
basis $\ket{\acc}$, $\ket{\rej}$.

A secure authentication scheme should, for any attack $\A$, produce
either the outcome $\ket{\rej}$ or the original state $\ket{\psi}$.
Of course, for a quantum system, a superposition of these would also
be acceptable.  Therefore, we let $\Pi (\ket{\psi})$ be the projector
onto the ``bad'' subspace containing states $\ket{\acc} \otimes
\ket{\phi}$, where $\ket{\phi}$ is orthogonal to $\ket{\psi}$.  We can
then define a secure quantum authentication scheme:

\begin{defn}
The encoding $\sigma_k (\ket{\psi})$ is a quantum authentication
scheme with error $\epsilon$ if, for all $\ket{\psi}$,
\begin{equation}
\frac{1}{|\K|} \sum_k \Tr \left[ \Pi(\ket{\psi}) \D(\sigma_{{\rm
Bob},k}(\ket{\psi}), k) \right] \leq \epsilon.
\end{equation}
\end{defn}

The following theorem was proved in previous work:~\cite{qas}

\begin{thm}
\label{thm:qasenc}
  A quantum authentication scheme with error $\epsilon$ is an
  encryption scheme with error at most $4\epsilon^{1/6}$.
\end{thm}

In fact, the family of quantum authentication schemes presented
previously~\cite{qas} all provided perfect encryption.  As a
consequence of theorem~\ref{thm:qasenc}, if we send a classical
message as a basis state for a quantum authentication scheme, it
is necessarily encrypted as well.  In fact, we can go even
further: it satisfies definition~\ref{def:unclone} for security
against cloning attacks.

\begin{thm}
\label{thm:qas}
A quantum authentication scheme with error $\epsilon$ is an
uncloneable encryption scheme with error at most $(15/2)
\epsilon^{1/6}$ (for small $\epsilon$).
\end{thm}

\proof

We can work with a purification of the original quantum authentication
scheme, so that Alice's encoding of a pure state $\ket{\psi}$ is again
a pure state for a given value $k$ of the classical key.  Then for a
particular value of the key, after Eve's attack and Bob's decoding, we
can write the state as
\begin{equation}
\ket{\acc} \ket{\psi} \ket{\phi_k} + \ket{\rej} \ket{\xi_k}
+ \ket{\acc} \ket{\eta_k},
\label{eq:decomp}
\end{equation}
where $\ket{\xi_k}$ and $\ket{\eta_k}$ are split between Bob and Eve,
with $\ket{\eta_k}$ orthogonal to $\ket{\psi}$.  $\ket{\phi_k}$ is
held by Eve.  Since this is a secure authentication scheme, $(\sum_k
\left\|\ket{\eta_k} \right\|^2)/|\K| \leq \epsilon$.  In general, the
various states depend on $\ket{\psi}$ as well as $k$ and $\A$.

When we send two possible classical messages $m$ and $m'$ with this
authentication scheme, we get
\begin{eqnarray}
m & \mapsto & \ket{\acc} \ket{m} \ket{\phi_k(m)} + \ket{\rej}
\ket{\xi_k(m)} + \ket{\acc} \ket{\eta_k(m)}, \\
m' & \mapsto & \ket{\acc} \ket{m'} \ket{\phi_k(m')} + \ket{\rej}
\ket{\xi_k(m')} + \ket{\acc} \ket{\eta_k(m')}.
\end{eqnarray}
However, since it is a quantum authentication scheme, Alice could have
sent $\ket{m} + \ket{m'}$ and it would have arrived safely:
\begin{equation}
(\ket{m} + \ket{m'}) \mapsto \ket{\acc} \Big(\ket{m} + \ket{m'}\Big)
\ket{\phi_k} + \ket{\rej} \ket{\xi_k} + \ket{\acc} \ket{\eta_k}.
\end{equation}
(Since $\ket{m} + \ket{m'}$ is not normalized, $\ket{\phi_k}$,
$\ket{\xi_k}$, and $\ket{\eta_k}$ are all bigger by a factor of
$\sqrt{2}$ than the corresponding terms in~(\ref{eq:decomp}).)  By
linearity,
\begin{eqnarray}
\ket{m} + \ket{m'} & \mapsto & \ket{\acc} \Big( \ket{m}
\ket{\phi_k(m)} + \ket{m'} \ket{\phi_k(m')} \Big) \\ & & + \ket{\rej}
\Big( \ket{\xi_k(m)} + \ket{\xi_k(m')} \Big) + \ket{\acc} \Big(
\ket{\eta_k(m)} + \ket{\eta_k(m')} \Big). \nonumber
\end{eqnarray}
The first term is the most interesting: if Eve's residual states
$\ket{\phi_k(m)}$ and $\ket{\phi_k(m')}$ are too different, they
become entangled with the message ket, and the state received by Bob
is no longer the superposition $\ket{m} + \ket{m'}$, but a mixture of
$\ket{m}$ and $\ket{m'}$.  In particular, we can write
\begin{eqnarray}
\ket{m} \ket{\phi_k(m)} + \ket{m'} \ket{\phi_k(m')} & = &
\Big(\ket{m} + \ket{m'} \Big) \otimes \Big(\ket{\phi_k(m)} +
\ket{\phi_k(m')}\Big)/2 \\
& & + \Big(\ket{m} - \ket{m'}\Big) \otimes \Big(\ket{\phi_k(m)} -
\ket{\phi_k(m')}\Big)/2. \nonumber
\end{eqnarray}
Thus
\begin{equation}
\ket{\eta_k} = \ket{\eta_k(m)} + \ket{\eta_k(m')} + \Big(\ket{m} -
 \ket{m'}\Big) \otimes \Big(\ket{\phi_k(m)} - \ket{\phi_k(m')}\Big)/2.
\label{eq:expansion}
\end{equation}
(Actually, some part of the first two terms could conceivably
contribute to $\ket{\phi_k}$, but that will only help the bound of
equation~(\ref{eq:residual}).)  Since $\ket{\eta_k}$,
$\ket{\eta_k(m)}$, and $\ket{\eta_k(m')}$ must all have small norm
for most $k$, so must the difference $\ket{\phi_k(m)} -
\ket{\phi_k(m')}$. In particular, for a fraction at least $1-1/q$
values of $k$, $\| \ket{\eta_k(m)} \|^2 \leq q \epsilon$, and
similarly for $\ket{\eta_k(m')}$ (with $q > 1$).  $\|\ket{\eta_k}
\|^2 \leq 2q \epsilon$ instead, because of normalization.  Thus,
for a fraction at least $1-3/q$, all three of the norms squared
are less than $q\epsilon$ (or $2q\epsilon$).
From~(\ref{eq:expansion}),
\begin{equation}
\Big\| \ket{\phi_k(m)} - \ket{\phi_k(m')} \Big\|/\sqrt{2} \leq
\big\|\ket{\eta_k}\big\| + \big\|\ket{\eta_k(m)}\big\| +
\big\|\ket{\eta_k(m')}\big\|.
\end{equation}
It follows that
\begin{equation}
\Big\| \ket{\phi_k(m)} - \ket{\phi_k(m')} \Big\| \leq 2 (1+\sqrt{2})
\sqrt{q \epsilon} \leq 5 \sqrt{q\epsilon}
\label{eq:residual}
\end{equation}
for the same fraction at least $1-3/q$ of the $k$'s.  At this point,
we are effectively done, since we have shown that Eve's residual
states are very similar for the two messages.  The remainder of the
proof is just massaging the formulas to get back to the correct form
for the definition of security.

\begin{claim}
\label{thm:trdist}
For those $k$s satisfying the above constraints,
\begin{equation}
D\Big(P(m)\rho_k (m), P(m')\rho_k (m')\Big) \leq q\epsilon + \sqrt{20
\sqrt{q \epsilon}} + 5 \sqrt{q\epsilon}
\label{eq:trdist}
\end{equation}
\end{claim}

The proof of this claim is fairly mechanical, and is given in
appendix~\ref{sec:trdist}.

Equation~(\ref{eq:trdist}) is valid for a fraction $1-3/q$ of the
possible values of $k$.  If we set $q = 2/(5\epsilon^{1/5})$ (which
will be greater than $1$ for small $\epsilon$), we find that
\begin{equation}
D\Big(P(m)\rho_k (m), P(m')\rho_k (m')\Big) \leq (1/2) \epsilon^{4/5}
+ 3.6 \epsilon^{1/5} + 3.2 \epsilon^{2/5} \leq \frac{15}{2}
\epsilon^{1/6}
\end{equation}
Note that $3/q$ is also less than $(15/2) \epsilon^{1/6}$,
completing the proof of the theorem.
\qed
\medskip

Theorem~\ref{thm:qas} provides a good way of constructing
uncloneable encryption schemes.  We need to come up with efficient
quantum authentication schemes, such as those given by Barnum {\it
et al}~\cite{qas}, and that gives us uncloneable encryption
schemes. However, it is not clear if this is the {\em only} way to
produce uncloneable encryption protocols.  Quantum authentication
schemes must authenticate the data in both the computational and
Fourier-transformed basis.  The proof of theorem~\ref{thm:qas}
suggests that authenticating in the Fourier basis is what produces
the uncloneability property, but there is no apparent reason we
need to also authenticate in the computational basis.  I therefore
conjecture that uncloneable encryption schemes exist which do not
authenticate the classical message and do not come from quantum
authentication schemes, but I am not aware of any examples.  If
true, the conjecture would imply that quantum authentication is a
strictly stronger property than uncloneability.

There is another point worth noting about theorem~\ref{thm:qas}: it is
an efficient reduction.  That is, given an attack against the
uncloneable encryption scheme, we can efficiently generate an attack
against the parent quantum authentication scheme.  This fact will be
critical in section~\ref{sec:compsec}, which discusses
computationally secure uncloneable encryption schemes.

\section{Uncloneable Encryption and QKD}
\label{sec:QKD}

A careful consideration of uncloneable encryption reveals that it is
closely related to another well-known quantum cryptographic task:
quantum key distribution.  In fact, any uncloneable encryption scheme
can be used to perform secure quantum key distribution.  In QKD, Alice
and Bob share authenticated classical channels and an insecure quantum
channel, and use just those resources to create a shared secret key.
Alice and Bob do not (generally) use any pre-existing secret key
beyond whatever is used in the classical authentication.

Given these same resources and an uncloneable encryption scheme, Alice
can perform QKD with the following protocol:
\begin{enumerate}
\item Alice generates random strings $k$ and $x$.
\item Alice sends the message $x$ to Bob using the uncloneable
encryption scheme with key $k$.
\item Bob announces (on the authenticated classical channel) that he
received the message.
\item Alice announces $k$ (again using the authenticated classical
channel).
\item Bob checks if the message is valid, and reports the result.  If
it is, Alice and Bob use $x$ as their new secret key.
\end{enumerate}
The properties of uncloneable encryption guarantee that this is a
secure QKD scheme: Eve gets the quantum state and then later learns
the key, but we know that her residual density matrices, conditioned
on Bob's accepting the transmission, are very similar.  Therefore, she
almost always (for most values of $k$) has little information about
the established key $x$.%
\footnote{In fact, most QKD schemes ask for an additional property:
that Alice and Bob should have the same value for the agreed-on key.
Since we do not require that an uncloneable encryption scheme
authenticate the classical message, we do not necessarily have this
property of QKD.  However, it can easily be guaranteed in one of two
ways: either classically authenticate $x$ before sending it with the
uncloneable encryption, or use an uncloneable encryption scheme
derived from quantum authentication, which will automatically
authenticate the classical message as well.}

In fact, this is a stronger security condition than what is
usually proved about QKD: most proofs only let Eve retain {\em
classical} information after QKD terminates, whereas here we are
letting Eve retain {\em quantum} information (although one
argument~\cite{mayers} contains a very similar statement imbedded
in the proof).  This is an important improvement, since it makes
it much easier to prove that the key generated via QKD can be used
in other cryptographic tasks.  For instance, a proof of the
security of the one-time pad goes as follows: We wish to show that
Eve cannot distinguish between two messages $a$ and $b$ given the
ciphertext $y$.  But this is equivalent to Eve distinguishing
between the two keys $x = a \oplus y$ and $x' = b \oplus y$, and
we know that the trace distance between Eve's residual density
matrices for these two cases is very small.  Therefore, Eve has
little chance of being able to distinguish the messages $a$ and
$b$.

While we can produce a QKD protocol from any uncloneable
encryption scheme, the converse is not necessarily true.  We shall
see in the next section that there is an uncloneable encryption
scheme that corresponds very closely to the BB84 QKD protocol, but
there are other QKD protocols which do not appear to have any
uncloneable encryption analogue.  For instance, QKD protocols with
two-way classical post-processing~\cite{twoway} are too
interactive to become uncloneable encryption schemes, but this is
perhaps not an important distinction since there are certainly
interactive quantum authentication schemes of the same form.

More interesting is the B92 QKD protocol~\cite{B92}.  In B92,
Alice sends one of two nonorthogonal states.  Each is part of a
particular basis for the Hilbert space.  Bob, when he receives the
transmission, randomly chooses one of the two bases to measure in
for each qubit. If he chooses the wrong basis, he has a chance of
getting a state orthogonal to the one Alice could have sent; in
that case, he {\em knows} he chose the wrong basis, and therefore
knows which basis Alice used.  In all other cases, the qubit is
discarded and does not influence the final key.  In BB84, many
qubits are also discarded, but this is simply an artifact of Bob's
initial ignorance of the basis, and is unnecessary if Bob has
quantum storage or, in the case of uncloneable encryption, prior
information about what the basis will be.  In contrast, in B92,
the basis choice determines the message sent, so Alice cannot ever
announce it and Bob cannot know it ahead of time.  Consequently,
discarding transmitted states is an intrinsic part of the
protocol.  For this reason, there does not appear to be an
uncloneable encryption analogue of B92.

We therefore have a situation where quantum authentication is slightly
stronger than uncloneable encryption, which is in turn slightly
stronger than quantum key distribution.  Nevertheless, the differences
are really quite small, meaning quantum authentication and quantum key
distribution are closely related.  This is rather surprising, given
that the tasks of authenticating quantum information and encrypting
classical information at first sight appear completely unrelated.

The connection between quantum authentication and quantum key
distribution helps us understand the conceptual structure of QKD.
The Shor-Preskill proof of security~\cite{SP} showed us that the
error correction and privacy amplification steps of QKD could be
seen as parts of a virtual quantum error correction procedure
taking place on some purification of the state.  Similarly, the
process of testing bit error rate can now be seen as coming from
quantum authentication: the error test in QKD comes from the
authentication test in the parent quantum authentication protocol.

\section{Uncloneable Encryption Without Entangled States}
\label{sec:prepandmeas}

In order to realize uncloneable encryption with near-future
technology, it is necessary to have a protocol which does not require
much in the way of coherent quantum manipulations, transmission, or
storage.  QKD is a good source of models; for instance, both BB84 and
B92 are ``prepare-and-measure'' protocols where Alice sends
unentangled qubits to Bob and Bob measures them immediately upon
receiving them, without having to store them at all.  We now wish to
find an uncloneable encryption protocol with this same structure.

One straightforward solution is simply to take BB84 and add a layer of
encryption.  That is, Alice encrypts her message with a one-time pad,
encodes it with an error-correcting code, and further encodes it as a set
of parity checks (coming from privacy amplification).  Then she sends
each bit of the resulting expanded message in one of two bases, and
intersperses at random a number of check bits.  The state of the check
bits, as well as the bases and other parameters, are determined by the
key, and Bob accepts the message only if the error rate on the check
bits is within an acceptable range.

The above proposal, when appropriately fleshed out, gives a secure
uncloneable encryption protocol.  However, the connection with
quantum authentication suggests more efficient ways of checking
for eavesdropping.  To start, I will construct a quantum
authentication protocol, and then use theorem~\ref{thm:qas} to
convert it to an uncloneable encryption protocol.  The resulting
protocol will involve a good deal of entanglement.  Then I will
use the technique of Shor and Preskill~\cite{SP} to convert this
to a ``prepare-and-measure'' protocol free of entanglement.  The
details of the construction require a certain amount of technical
background beyond the scope of this paper, so I postpone the
derivation and proof of security to appendix~\ref{sec:pmconstr}.
In this section, I simply present the resulting
``prepare-and-measure'' protocol and discuss some extensions and
applications to QKD.

The uncloneable encryption scheme will be designed to work through a
noisy channel and will depend on a choice of two classical linear
codes.  $C_1$ will be used to correct bit flip errors in the data.
Let $\delta$ be the average of the rates of bit flip and phase errors
introduced by the noisy channel; since the qubits transmitted will be
sent in one of two bases, $\delta$ will be the actual rate of errors
in the transmitted message in the absence of an eavesdropper.  $C_1$
will encode $K$ bits in $N$ bits and will have distance $2\delta N$;
that is, it will correct a fraction $\delta$ errors.  By the
Gilbert-Varshamov bound, such a code exists (for large $N$) with $K/N
\geq 1 - h(2\delta)$, where $h(x) = - x \log_2 x - (1-x) \log_2 (1-x)$
is the binary entropy function.  $C_1$ is defined by $N-K$ parity
checks, which can be encapsulated in a parity check matrix $H$.  The
set of vectors $\bf{v}$ with a particular value for $H \bf{v}$ is
called a coset of $C_1$; the error-correcting code is generally
considered to be the coset with $H \bf{v} = \bf{0}$, but all the other
cosets have the same error-correcting properties.  We have, for a
particular $C_1$, some standard decoding algorithm $M$ which maps
elements of one coset into another.  Since the code has distance $2
\delta N$, different errors of weight less than $\delta N$ take us to
different cosets, and $M$ can therefore be chosen to correct any such
error.  In order to have a practical protocol, we actually need to
choose a carefully-constructed $C_1$ so that $M$ can be implemented
efficiently.  This is a challenging task, and the focus of much of
classical coding theory, and I will for this paper simply assume that
we have an efficient decoding algorithm.

The second classical code, $C_2$, is used to perform what, in the
context of BB84, would be called privacy amplification.  Eve might
choose to measure only a few of the transmitted qubits, in which
case she is unlikely to be detected by Bob; if she gets lucky in
choosing bases, we need to be sure she still gets little
information about the data (after she learns the key).  $C_2$ must
correct a slightly larger fraction of errors than $C_1$, so we
will give it distance $2 (\delta + \eta) N$.  $C_2$ encodes $K'$
bits in $N$ bits and satisfies $C_2^\perp \subset C_1$ (where
$C_2^\perp$ is the standard classical dual code containing all
vectors orthogonal to $C_2$ in the usual binary inner product). We
can choose a $C_2$ with these properties and $K'/N \geq 1 -
h(2\delta + 2\eta)$~\cite{CS}.  We will not require
that $C_2$ have an efficient decoding algorithm.%
\footnote{Actually, we only require that $C_2$ correct errors
modulo $C_1^\perp$ (i.e., errors which sum to elements of
$C_1^\perp$ are considered equivalent) due to the existence of
degenerate quantum codes.  This is potentially important, as it is
unclear whether it is possible to choose an efficiently decodeable
$C_1$ such that $C_2$ has good minimum distance, but it is much
easier to do so with the looser requirement on $C_2$.}

The actual uncloneable encryption protocol will encrypt a message $m$
of length $n$ bits.  There will be a security parameter $s$ and we
will choose $K + K' - N = n+s$.  The protocol will transmit $N$ qubits
and will use a classical key consisting of four parts: $(k, e, c_1,
b)$.  All are chosen uniformly at random; their lengths are described
below.

\begin{enumerate}

\item Divide the $n$ input bits into $r$ groups of size $s$.  View
each group of $s$ bits as an element of the finite field $GF(2^s)$.
$k$ is a string of size $s$, and we can also view it as an element of
$GF(2^s)$.

\item The $r$ resulting registers $m_0, \ldots, m_{r-1}$ can be viewed
as the first $r$ coefficients of a degree $r$ polynomial $f$.  The
final coefficient $m_r$, the constant term, is chosen so that $f(k) =
0$.  That is, $\sum_i m_i k^{r-i} = 0$ (in $GF(2^s)$).

\item Alice XORs the string $(m_0, \ldots, m_r)$ with the $(n+s)$-bit
string $e$, producing a new classical string $y$ of length $n+s$ bits.

\item Alice considers the particular coset of the classical
error-correcting code $C_1$ given by the syndrome $c_1$; that is, she
considers the coset satisfying $H{\bf v} = {\bf c_1}$.  (The length of
$c_1$ is equal to the number of parity checks, namely $N-K$.)

\item Within that coset of $C_1$ are various cosets of $C_2^\perp$
(which, recall, is a subset of $C_1$).  In fact, there are exactly
$2^{n+s}$ of them, and they can be distinguished by parity checks
which are elements of the code $C_2$ but are not in $C_1^\perp$
(since all elements of a given coset of $C_1$ have the same value
for parity checks from $C_1^\perp$).  There is thus a
correspondence between the cosets $C_1/C_2^\perp$ and strings $y$
(from step $3$).  Alice selects the coset corresponding to $y$ and
then picks a random $N$-bit string $z$ within that coset.

\item Alice transmits $N$ qubits as follows: When the $i$th bit of $b$
is $0$, Alice transmits the $i$th bit of $z$ in the $Z$ basis
($\ket{0}$, $\ket{1}$).  When the $i$th bit of $b$ is $1$, Alice
transmits the $i$th bit of $z$ in the $X$ basis ($\ket{0}+\ket{1}$,
$\ket{0}-\ket{1}$).

\end{enumerate}
To decode, Bob simply reverses this series of actions:
\begin{enumerate}

\item Bob receives $N$ qubits.  When the $i$th bit of $b$ is $0$, he
measures the $i$th qubit in the $Z$ basis; when the $i$th bit of $b$
is $1$, he measures the $i$th qubit in the $X$ basis.  He now has an
$N$-bit classical string $z$.

\item Bob calculates the parity checks of the classical code
$C_1^\perp$.  If they are not equal to the string $c_1$, there are
errors in the state, which he corrects using the standard decoding map
$M$.

\item Bob evaluates the parity checks of $C_2/C_1^\perp$, producing a
$(n+s)$-bit string $y$.  (This step is effectively the privacy
amplification step in BB84.)

\item Bob XORs $y$ with the $(n+s)$-bit string $e$, producing a new
string $(m_0, \ldots, m_r)$.

\item Bob has now received the $n$-bit message $(m_0, \ldots,
m_{r-1})$.  To check whether he accepts this or not (that is, to
detect eavesdropping by Eve), he considers the $s$-bit registers $m_0,
\ldots, m_r$ as elements of $GF(2^s)$ and coefficients of a degree $r$
polynomial $f$ over $GF(2^s)$.  He evaluates $f(k)$ and accepts the
message only if $f(k) = 0$.

\end{enumerate}

The shared classical key, as noted above, consists of $(k, e, c_1,
b)$.  $k$ is an $s$-bit string, $e$ is $n+s$ bits long, $c_1$ is $N-K
\leq h(2\delta) N$ bits, and $b$ is $N$ bits long.  $s$ and $\eta
\sqrt{N}$ are the security parameters: the security $\epsilon$ of the
uncloneable encryption protocol is exponentially small when the two
security parameters are large.  In particular, $\eta$ can go to $0$ as
$n$ becomes large without severely impacting the security of the
protocol.  All in all, then, Alice and Bob use $n+2s + N - K$ bits of
key.  In the limit of a perfect channel (i.e., $\delta = 0$), they use
$n+2s + (n+s)/[1-h(2\eta)]$ bits of key, which asymptotes to $2n+3s$
as $n$ becomes large and $\eta$ becomes small.  In addition, Alice
chooses $N - K'$ random bits which are not part of the key.  Bob need
not know the values of these bits ahead of time.  On the other hand,
these random bits must never be revealed to Eve, even after Bob has
successfully received the transmission (when, by the uncloneability
property, it is safe for Eve to learn the key).  Otherwise the
benefits of the privacy amplification step would be eliminated, and
Eve could possibly learn a few bits of information about the message
(though only if she also knows $e$).

For the basic protocol, we assume the full key is used for just a
single message, which would suggest uncloneable encryption uses up
more than twice as much key as the one-time pad.  However, we can
partially lift this restriction by encrypting the whole message.
Instead of the secret keys $e$ and $c_1$, we use a key $e'$ of length
$N$ to encrypt the final string $z$ before encoding it as quantum
states.  Because of this encoding, the state being transmitted is the
identity density matrix, and therefore Eve, even if she knows the
original message $m$, has no information about the keys $k$ and $b$.
(She does learn something about $e'$, though.)  We can therefore reuse
the keys $k$ and $b$ in later messages and the system remains secure,
so long as we use a new $e'$ for each message.  This gives us a
reusable key $(k,b)$ of size $s + N$ and a one-time key $e'$ of length
$N$ which must be refreshed for each new message.  In the limit of
long messages and low channel error rate, the key expended per
uncloneably encrypted message is asymptotically the same as that used
by a message encrypted with the one-time pad.

In a very real sense, the reusable key provides the uncloneability
property and the one-time key provides the encryption.  If Eve, after
seeing many messages, ever learns the reusable key $(k, b)$, by the
uncloneability property, she still has no information about past
messages.  Future messages remain encrypted with the one-time pad,
since they use new values of $e'$, but Eve can easily copy these
messages, and if she eventually learns $e'$ for one of these later
messages, she can read the message.  On the other hand, if after
transmission, Eve ever learns $e'$ for a particular message, she again
has no information about that message (by uncloneability), but she
might be able to learn information about the values of $k$ and/or $b$
used for that message.  This would imperil unclonability of future
messages.

The above protocol has a ``prepare-and-measure'' form and
therefore might be suited to experimental realization sometime in
the near future.  However, it is somewhat more challenging than
the closely related BB84 QKD protocol.  In particular, in BB84, it
is harmless to discard any qubits which are not received by Bob
for whatever reason. In contrast, for uncloneable encryption,
discarded bits cut away at the error-correcting code protecting
the data.  In fact, they act as erasure errors and should be
treated as such.  A quantum error-correcting code can only
tolerate a certain proportion of erasures (half if there are no
other errors in the system), and therefore uncloneable encryption
will only work if the rate of photon absorption is not too high.
Therefore, uncloneable encryption requires high efficiency
single-photon sources and detectors, which are useful for QKD but
not required.  This restriction can be seen as a cost of going to
a non-interactive uncloneable encryption protocol rather than an
interactive QKD protocol.  One possible solution is to use a
squeezed-state cryptographic scheme~\cite{cont}, which largely
avoids this problem.

Alternatively, it is straightforward to use the above protocol for
QKD.  Alice simply sends a string of bits in some series of bases
$b$, Bob measures in whichever bases he likes, and they keep only
those qubits which are received and for which their bases agree.
Alice announces $b$ along with randomly chosen values for the
other parts of the key, and then does the same decoding procedure
as Bob, ensuring that they agree on a final secret key.  There are
a few advantages to doing this over the usual approach to QKD.
First, we get the strong security condition described in
section~\ref{sec:QKD}.  Second, because of the connection with
quantum authentication, it is clear how to create protocols like
the one from this section which use a much more efficient test for
eavesdropping than the usual prescription for BB84 (which reveals
a substantial fraction of the originally transmitted bits to
compare error rates).  Third, Alice and Bob could take their bases
from a pseudorandom sequence generated by a short shared key, as
discussed in the next section.  Then, instead of discarding half
of all bits received, Bob can be sure he measures every qubit in
the same basis Alice used.  The result is no longer
unconditionally secure, but they instead have unconditional {\em
forward} secrecy: provided Eve is faced with a computational
limitation during the initial transmission, she cannot ever learn
the established key, even after her computational bound is lifted.
(This result will be shown in the following section.)
Unconditional forward secrecy is also available in some classical
protocols, but those require an external source of randomness and
a potentially unrealistic temporary {\em memory} bound on the
adversary~\cite{maurer,rabin1,rabin2}.  Efficient QKD
protocols~\cite{efficient} (in which Alice and Bob use the two
bases with unequal probabilities) can also reduce the number of
qubits discarded.  The efficient QKD protocols do not require even
a temporary computational assumption, but also do not completely
eliminate wasted transmissions.

Note that it is insecure to combine the reusable key refinement with
QKD.  Since the one-time key $e'$ must be announced when performing
QKD, Eve learns it and can therefore learn information about the
supposedly reusable key $(k,b)$.  She could then use this information
to break later QKD protocols which attempt to reuse $(k,b)$.

\section{Computational Security}
\label{sec:compsec}

The key requirements of uncloneable encryption are not immense
(roughly $2n$ for long messages), but are still more than is
desireable in a truly non-interactive setting (where QKD is not
available to produce more key).  In classical cryptography, we
frequently use a computational assumption to encrypt long messages
with a short key.  Does uncloneable encryption still work if the key
is not truly random, but is instead a {\em pseudorandom} sequence
generated from a much shorter secret key shared by Alice and Bob?

A similar question arises in the context of QKD.  Alice must make a
lot of random choices when preparing the qubits to send to Bob.
Generating truly random numbers can be a difficult task.  If she
instead generates a long pseudorandom sequence, what does that do to
the security of QKD?

In both cases, the answer is that Eve still cannot learn the secret
message, provided she has no quantum algorithm to break the
pseudorandom sequence.  Furthermore, even if she can {\em eventually}
break the computational assumption, it will do her no good: in order
to defeat Alice and Bob, Eve must break the pseudorandom sequence
before Bob receives the quantum transmission from Alice.  Intuitively,
this makes a lot of sense: unless she can defeat the scheme during
transmission, the uncloneability property holds, preventing her from
copying the message down to work on it later.  (Note, though, that the
values of the bits Alice transmits in QKD, including any lost to
privacy amplification, should be truly random or there is no hope of
long-term security once the computational restriction is lifted.)

I will prove this in the case of an uncloneable encryption scheme
$\Sc$ derived from quantum authentication schemes under
theorem~\ref{thm:qas}.  In particular, this holds for the uncloneable
encryption scheme presented in section~\ref{sec:prepandmeas}.  The
definition of a pseudorandom sequence is one which a
computationally-bounded Eve cannot distinguish from a truly random
sequence.

\begin{thm}
\label{thm:comp}
If Eve can break uncloneable encryption scheme $\Sc$ (which is derived
from a quantum authentication scheme) with a pseudorandom key (from
oracle $\K$) using an attack of low complexity during transmission,
then she has an efficient quantum algorithm that can distinguish $\K$
from a truly random sequence.
\end{thm}

\proof

The proof is straightforward.  Eve is given a string $k$, and wishes
to determine if it is a pseudorandom string generated by $\K$ or a
truly random sequence.  Now, she has an attack $\A$ which breaks the
uncloneable encryption scheme $\Sc$ when it uses a pseudorandom key.
That is, there are two messages $m$, $m'$ for which the probabilities
$P(m)$, $P(m')$ of being accepted are not both small, and for which
Eve's residual density matrices $\rho_k(m)$, $\rho_k(m')$ are
substantially different.

By theorem~\ref{thm:qas}, she therefore can efficiently generate an
attack $\A'$ against the quantum authentication scheme $\Sc$ with key
drawn from $\K$.  That is, there is some input state $\ket{\psi}$
(which, by the proof of theorem~\ref{thm:qas}, we know can be chosen
to be $\ket{m} + \ket{m'}$) for which the output of the authentication
scheme with attack $\A'$ has a large component which is accepted but
is orthogonal to $\ket{\psi}$.

To break the pseudorandom sequence, Eve therefore creates a simulated
Alice sending messages to a simulated Bob using the quantum
authentication scheme $\Sc$ and key $k$.  The pretend Alice repeatedly
sends the message $\ket{\psi}$ to pretend Bob using this key.  Each
time, Eve performs the attack $\A$ and measures the state received by
Bob in an orthonormal basis including $\ket{\psi}$.

We know the quantum authentication scheme is secure when used with a
truly random key.  Therefore, if the key is truly random, Eve will
essentially always find that the simulated Bob either rejects the
message or receives the state $\ket{\psi}$.  On the other hand, the
attack $\A'$ breaks the protocol when the key is pseudorandom, so when
$k$ is generated by $\K$, Eve will occasionally find that the
simulated Bob accepts a state which she measures to be orthogonal to
$\ket{\psi}$.  Therefore, if Eve ever gets such an outcome, she
concludes $k$ is pseudorandom; otherwise, she decides it is random.
\qed
\medskip

Note that, while the protocol given in section~\ref{sec:prepandmeas}
does not come directly from a quantum authentication protocol, it
still is derived from one indirectly.  In particular, combining
theorem~\ref{thm:qas} and appendix~\ref{sec:pmconstr}, we produce an
efficient reduction to a quantum authentication scheme.  Thus,
theorem~\ref{thm:comp} also holds for that protocol and similar
prepare-and-measure BB84 protocols.

\section{Open Questions}
\label{sec:open}

One serious drawback of theorem~\ref{thm:comp} is that it only proves
security when the pseudorandom sequence is secure against {\em
quantum} attacks.  Intuitively, we should expect that if we use a
``prepare-and-measure'' protocol, such as that in
section~\ref{sec:prepandmeas}, and Eve can make only attacks against
individual photons, she should not be able to copy the state, even if
the pseudorandom sequence is only secure against classical attacks.
The whole protocol and attack can be simulated classically, so Eve is
not sneaking in any additional computational power with the attack,
and therefore should not be able to copy the state.  It would be
extremely useful to prove that under these circumstances, the
uncloneability property still holds.  That would allow us, for
instance, to perform ``prepare-and-measure'' uncloneable encryption
for message $m$ today based on an RSA-protected key, and still have
information-theoretic security for $m$ in the distant future once
quantum computers able to break RSA become available.

A useful practical improvement for the ``prepare-and-measure''
protocol from section~\ref{sec:prepandmeas} would be to give it more
tolerance to channel noise.  In particular, the requirement that $C_1$
and $C_2$ have good minimum distance is rather strict.  Perhaps this
can be improved to allow $C_1$ and $C_2$ to correct general errors,
rather than worst-case errors, with error rates $\delta$ and $\delta +
\eta$.

It would also be interesting to understand the connection, if any,
between uncloneability and key reuse for quantum authentication.
Recent work~\cite{HOreuse,HLMreuse} has shown that much (but not
all) of the key for quantum authentication can be reused, provided
the authentication test indicates no tampering.  There is clearly
some similarity with the fact that quantum authentication schemes
provide uncloneable encryption, but neither result is strictly
stronger or weaker than the other: The fact that part of the key
can be reused indicates that it can safely be exposed after the
message is successfully received, but uncloneability allows {\em
all} of the key to become public.  On the other hand, key can be
reused in ways other than publishing it to Eve, and uncloneability
does not imply the security of those other applications.

One might also wish to give the uncloneability property to other types
of classical protocols.  For instance, one can make a simple
uncloneable secret sharing scheme with two shares.  To do this, take a
classical secret string $m$ and share it as $(a_0, a_1)$ for random
string $a_0$, and $a_1 = m \oplus a_0$.  Both shares are now needed to
reconstruct $m$.  We can make this scheme uncloneable by using an
uncloneable encryption scheme $\Sc_k$.  Encrypt $a_i$ with $\Sc$ using
key $k_i$.  The first share of the new uncloneable scheme is
$(\Sc_{k_0}(a_0), k_1)$, and the second share is $(\Sc_{k_1}(a_1),
k_0)$.  That is, each share contains an encrypted share of the
original classical scheme, plus the key needed to decode the other
share.  Someone with only one share of this new scheme cannot copy it
without being detected, although of course anyone with both shares can
read them and copy them without difficulty.  To create more complex
secret-sharing schemes, we need a good definition of uncloneable
secret sharing in general.  What shares is the adversary allowed
access to, and when (and with what shares) do the users check for
intrusion?

One difficulty with such generalizations is that it is unclear to what
extent the name ``uncloneable encryption'' is really deserved.  I have
not shown that a message protected by uncloneable encryption cannot be
copied --- only that Eve cannot copy it without being detected.  Is it
possible for Eve to create two states, neither of which will pass
Bob's test, but which can each be used (in conjunction with the secret
key) to extract a good deal of information about the message?  Or can
one instead prove bounds, for instance, on the sum of the information
content of the various purported copies?

Another interesting open question is to better understand the
relationships of various cryptographic tasks.  Is quantum
authentication exactly equivalent to quantum key distribution in some
sense, or is there a real distinction between the two?  Also, are
other cryptographic tasks for quantum information related to
cryptographic tasks for classical data?  There may well be a rich
structure of interconnections between quantum and classical protocols
waiting to be uncovered.

Finally, the task of uncloneable encryption is not really
conceivable in a purely classical context.  Are there other useful
tasks waiting to be discovered which simply make no sense for a
classical protocol yet are achievable with the aid of quantum
information?

\section*{Acknowledgements}

I would like to thank Claude Cr\'epeau, Hoi-Kwong Lo, Dominic Mayers,
and Adam Smith for helpful discussions.  This work was performed in
part while the author was a Clay Long-Term CMI Prize Fellow.

\appendix

\section{Proof of Claim~\ref{thm:trdist}}
\label{sec:trdist}

We wish to show equation~(\ref{eq:trdist}):
\begin{equation}
D\Big(P(m)\rho_k (m), P(m')\rho_k (m')\Big) \leq q\epsilon + \sqrt{20
\sqrt{q\epsilon}} + 5 \sqrt{q\epsilon}.
\end{equation}

The residual density matrices $\rho_k$ for Eve in the case she is
not caught are
\begin{eqnarray}
\rho_k (m) & = & \frac{\ketbra{\phi_k(m)} + \Tr_{\rm Bob}
\ketbra{\eta_k(m)}}{N(m)^2 + \|\ket{\eta_k(m)}\|^2}, \\
\rho_k (m') & = & \frac{\ketbra{\phi_k(m')} + \Tr_{\rm Bob}
\ketbra{\eta_k(m')}}{N(m')^2 + \|\ket{\eta_k(m')}\|^2},
\end{eqnarray}
where $N(m) = \|\ket{\phi_k(m)}\|$ and $N(m') = \|\ket{\phi_k(m')}\|$.

Then, by the triangle inequality,
\begin{eqnarray}
D\Big(P(m) \rho_k (m), P(m') \rho_k (m')\Big) & \leq &
D\Big(P(m) \rho_k (m), \ketbra{\phi_k(m)}\Big) \label{eq:DFull}\\
& & + D\Big(\ketbra{\phi_k(m)}, \ketbra{\phi_k(m')} \Big)
\nonumber \\
& & + D\Big(P(m') \rho_k (m'), \ketbra{\phi_k(m')}\Big). \nonumber
\end{eqnarray}
Now,
\begin{eqnarray}
D\Big(P(m) \rho_k (m), \ketbra{\phi_k(m)}\Big) & = &
\frac{1}{2} \Tr \big|  \Tr_{\rm Bob} \ketbra{\eta_k(m)}\big| \\
& \leq & q\epsilon/2, \label{eq:DOne}
\end{eqnarray}
and similarly,
\begin{equation}
D\Big(P(m') \rho_k (m'), \ketbra{\phi_k(m')}\Big) \leq q\epsilon/2.
\label{eq:DTwo}
\end{equation}

We need, therefore, only to bound
\begin{equation}
D\Big(\ketbra{\phi_k(m)}, \ketbra{\phi_k(m')} \Big). \nonumber
\end{equation}
First, note that
\begin{eqnarray}
\lefteqn{D\Big(\ketbra{\phi_k(m)},\ketbra{\phi_k(m')}\Big) \leq}
\qquad \nonumber \\
& & N(m)^2 D\left(\frac{1}{N(m)^2}\ketbra{\phi_k(m)},
\frac{1}{N(m')^2}\ketbra{\phi_k(m')}\right) \nonumber \\
& & + D\left(\frac{N(m)^2}{N(m')^2}\ketbra{\phi_k(m')},
\ketbra{\phi_k(m')}\right) \\
& = & N(m)^2 D\left(\frac{1}{N(m)^2}\ketbra{\phi_k(m)},
\frac{1}{N(m')^2}\ketbra{\phi_k(m')}\right) \nonumber \\
& & + \frac{1}{2} \left| N(m)^2 - N(m')^2 \right|.
\label{eq:DThreea}
\end{eqnarray}

Now,
\begin{eqnarray}
\lefteqn{\braket{\phi_k(m')}{\phi_k(m')} -
\braket{\phi_k(m')}{\phi_k(m)} =} \qquad \nonumber \\
& & N(m')^2 - N(m)N(m') F\left(\frac{1}{N(m)}\ket{\phi_k(m)},
\frac{1}{N(m')}\ket{\phi_k(m')}\right) \\
& \leq & \big\| \ket{\phi_k(m')} \big\| \cdot \Big\| \ket{\phi_k(m)} -
\ket{\phi_k (m')} \Big\| \\
& \leq & N(m') 5 \sqrt{q \epsilon},
\end{eqnarray}
where the last line follows from equation~(\ref{eq:residual}).  That
is,
\begin{eqnarray}
F\left(\frac{1}{N(m)} \ket{\phi_k(m)}, \frac{1}{N(m')}
 \ket{\phi_k(m')}\right)
& \geq & \frac{N(m') - 5 \sqrt{q \epsilon}}{N(m)} \\
& = & 1 - \frac{(N(m) - N(m')) + 5 \sqrt{q \epsilon}}{N(m)}.
\end{eqnarray}
By the triangle inequality and equation~(\ref{eq:residual}),
\begin{equation}
N(m) - N(m') \leq 5 \sqrt{q \epsilon},
\label{eq:normdif}
\end{equation}
so
\begin{equation}
F\left(\frac{1}{N(m)} \ket{\phi_k(m)}, \frac{1}{N(m')}
\ket{\phi_k(m')}\right)
\geq 1 - \frac{10 \sqrt{q \epsilon}}{N(m)}.
\end{equation}
Thus,
\begin{eqnarray}
D\left(\frac{1}{N(m)^2}\ketbra{\phi_k(m)},
\frac{1}{N(m')^2}\ketbra{\phi_k(m')} \right) & \leq & \sqrt{1-F^2} \\
& \leq & \sqrt{\frac{20 \sqrt{q \epsilon}}{N(m)}}.
\label{eq:DThree}
\end{eqnarray}

Putting together equations~(\ref{eq:DFull}), (\ref{eq:DOne}),
(\ref{eq:DTwo}), (\ref{eq:DThreea}), and (\ref{eq:DThree}), we get
\begin{eqnarray}
D\Big(P(m)\rho_k(m), P(m')\rho_k(m')\Big) & \leq & q\epsilon +
N(m) \sqrt{20 N(m) \sqrt{q \epsilon}} \nonumber \\
& & + \frac{1}{2} \left| N(m)^2 - N(m')^2 \right| \\
& \leq & q\epsilon + \sqrt{20 \sqrt{q \epsilon}} +
5\sqrt{q\epsilon},
\end{eqnarray}
where we have used $N(m) \leq 1$ and equation~(\ref{eq:normdif}) to
bound $|N(m)^2 - N(m')^2|$.

\section{Constructing a Prepare-and-Measure Protocol}
\label{sec:pmconstr}

We wish to construct a ``prepare-and-measure'' protocol by
starting with a quantum authentication protocol of an appropriate
form.  The easiest way to create an efficient quantum
authentication protocol is to use the technique of Barnum {\it et
al}~\cite{qas}: create a set of stabilizer codes with the right
properties --- in the terminology of Barnum {\it et al}, they form
a ``Purity Testing Code.''  Then this will give us a quantum
authentication protocol.

However, there is a complication.  We wish to end up with a
protocol that, like BB84, works even through a noisy channel.  One
obvious way to do this would be to encode the quantum
authentication protocol with a quantum error-correcting code, but
this would destroy the prepare-and-measure structure we wish to
have for the final protocol. Instead, we will devise a quantum
authentication protocol with the additional ability to correct
errors.  An examination of~\cite{qas} reveals that this does not
require modification of the definition of quantum authentication;
we simply add the property that the transmission (almost always)
is accepted when the data is sent through some particular channel
$\C$.

We will again construct such a protocol from a family of stabilizer
codes.  Recall that a stabilizer quantum error-correcting code is an
abelian subgroup of the Pauli group generated by tensor products of
the Pauli matrices
\begin{equation}
X = \pmatrix{0 & 1 \cr 1 & 0}, \quad
Y = \pmatrix{0 & -i \cr i & \ 0}, \quad
Z = \pmatrix{1 & \ 0 \cr 0 & -1}.
\end{equation}
A stabilizer code $Q$ detects a Pauli error unless that error is in
the set $Q^\perp - Q$, where $Q^\perp$ is the set of Pauli operations
that commute with every element of the stabilizer.  A stabilizer code
corrects a set of Pauli errors $\E$ if it detects the product of any
two elements of $\E$.  We can also talk about a stabilizer code which
corrects a set $\E$ and detects a larger set $\F \supseteq \E$.  This
occurs if the code detects all operators which are the product of an
element from $\E$ and an element from $\F$.  (When this is true, the
code can distinguish elements of $\F$ from elements of $\E$ and can
distinguish elements of $\E$ from each other, but cannot necessarily
tell elements of $\F$ apart.)  For more detailed background on
stabilizer codes, see~\cite{stabintro}.

\begin{defn}
\label{def:ptcqecc}
Let $\{Q_k\}$ be a family of stabilizer codes.  Suppose the code $Q_k$
corrects the set $\E_k$ and detects the set $\F_k$ (for some decoding
algorithm, which will vary with $k$).  Let $\C$ be a Pauli channel
(i.e., it produces Pauli errors with various probabilities).  Then the
set $\{Q_k\}$ is a purity testing code with error $\epsilon$ which
corrects the channel $\C$ if the following conditions hold true:
\begin{enumerate}
\item For a fraction at least $1-\epsilon$ of the possible values of
$k$, $\E_k$ contains a typical set of Pauli errors produced by $\C$
(one that occurs with probability at least $1-\epsilon$).
\item For {\em any} Pauli error $E$ (not the identity), $E \in \F_k$
for a fraction at least $1-\epsilon$ of the possible values of $k$.
\end{enumerate}
\end{defn}

The first condition is straightforward --- it simply means that
most codes in the family should correct errors from the channel.
The second condition is a little more slippery.  It says that most
of the codes will detect {\em any} particular Pauli error, despite
the general error correction that is going on.  A slight
modification of the theorems in~\cite{qas} shows that a family of
codes satisfying definition~\ref{def:ptcqecc} can be used to
construct a quantum authentication protocol which also corrects
the channel $\C$.

To create such a family of codes, I will use a concatenated structure.
The inner layer will be a classical authentication protocol, giving a
family of modest size.  This part will serve to detect errors which
are not corrected.  Then will come a fixed quantum error-correcting
code (or perhaps a family of quantum error-correcting codes), which
are designed to correct a channel related to $\C$, which I assume is
memoryless.  Finally, for the outside layer, we will perform Hadamard
transforms on some of the qubits; the set of qubits transformed is
determined by part of $k$.  This will serve to mix phase and bit flip
errors, allowing the classical authentication protocol to detect the
phase errors as well as the bit flip errors.

In principle, any classical authentication protocol should suffice for
the inner layer of encoding.  However, the proof techniques used
require it to be describable in the stabilizer language, which limits
us somewhat.  On the other hand, the condition for
definition~\ref{def:ptcqecc} is somewhat weaker than an actual
classical authentication protocol.  For instance, we can use the
following encoding: Suppose we have an $n$ bit message $m$, with
$n=rs$.
\begin{enumerate}
\item Divide the $n$ input bits into $r$ groups of size $s$.  View
each group of $s$ bits as an element of the finite field $GF(2^s)$.
Let $k$ be a secret string of size $s$, also viewed as an element of
$GF(2^s)$.
\item The $r$ resulting registers $m_0, \ldots, m_{r-1}$ can be viewed
as the first $r$ coefficients of a degree $r$ polynomial $f(z)$.  The
final coefficient $m_r$, the constant term, will be chosen so that
$f(k) = 0$.  That is, $\sum_i m_i k^{r-i} = 0$.
\item Alice sends $(m_0, \ldots, m_r)$ to Bob, who accepts the message
if he receives a list of registers defining a polynomial $f'$ with
$f'(k) = 0$.
\end{enumerate}
The property we will need for definition~\ref{def:ptcqecc} is that, if
Eve adds any nonzero vector to the transmission, then for a fraction
$1-\epsilon$ of the values of $k$, the resulting string fails the
test.  To see this, note that the test will be passed only if Eve's
vector corresponds to a degree-$r$ polynomial $g$ with $g(k)=0$.
However, a polynomial of degree $r$ can have at most $r$ zeros, so
whatever polynomial $g$ Eve picks, it can pass the test for at most a
fraction $r/2^s = n/(s2^s)$ of the possible values of $k$.  For
suitable $s$, we can easily make this very small.

Note that this is not necessarily a complete classical authentication
protocol: $GF(2^s)$ is not algebraically complete, so whatever
polynomial Alice sends might have fewer than $r$ zeros.  In that case,
Eve could safely replace it by a different polynomial with the same
(or a larger) set of zeros.  Nevertheless, this protocol will suffice
to give a quantum authentication protocol in conjunction with the
other processing we apply, namely encryption.

We now have a way of encoding an $n$-bit classical message as $n+s$
bits; we can extend this linearly to encode superpositions as well, so
$n$ qubits are encoded as $n+s$ qubits.  We then take a quantum
error-correcting code which encodes $n+s$ qubits in $N$ qubits, and
finally, based on an $N$-bit classical string $b$, perform Hadamard
transforms on many of the qubits.  If the $i$th bit of $b$ is a $0$,
we leave the $i$th qubit of the encoded message alone; if the $i$th
bit of $b$ is $1$, we perform the Hadamard on the $i$th qubit.  The
result is a purity testing code which also corrects errors.

The quantum error-correcting code must have three properties.  First,
it must usually correct the channel $\C$.  If we choose a random $b$,
we are effectively swapping the $X$ (bit flip) and $Z$ (phase flip)
error rates for half the qubits transmitted.  In other words, we have
transformed to a new channel $\C'$ which is symmetrized between bit
and phase flip errors.  (The $Y$ error, which does both a bit and
phase flip, could have a different error rate.)  Our error-correcting
code should correct this channel.

Second, in order to perform the final Shor-Preskill step of our
reduction to a ``prepare-and-measure'' protocol, we will need a
CSS code~\cite{CS,Steane}.  Let $\delta$ be the bit error rate of
the symmetrized channel $\C'$ (i.e., the combined $X$ and $Y$
error rates, or equivalently the combined $Y$ and $Z$ error
rates).  Then, in particular, we want a CSS code that corrects
$\delta N$ bit flip errors and $\delta N$ phase errors.  Actually,
we should pick $\delta$ to be slightly larger than the true error
rate so the code can also correct statistical fluctuations from
the average error rate.

Third, we need our full concatenated construction to act as a purity
testing protocol.  That is, suppose Eve presents us with an arbitrary
Pauli operation $E$.  We wish either for this error to be corrected by
the quantum code or detected by the authentication procedure.  To
analyze this, we will see how $E$ is treated by each level of the
decoding procedure.  In the outer layer, we perform Hadamard
transforms according to the bit string $b$.  Whenever we perform a
Hadamard, we convert an $X$ to a $Z$ and vice-versa.  This gives us a
new Pauli operation $E'$.  In most cases, $E'$ will have a similar
number of $X$s and $Z$s (the number of $Y$s may be different).  The
total values $\delta_X$ (the total fraction of $X$s or $Y$s) and
$\delta_Z$ (the total fraction of $Y$s or $Z$s) will therefore be very
similar on average.  If $\delta_X \leq \delta$, the quantum
error-correcting code will correct the bit flip part of the errors; if
$\delta_Z \leq \delta$, the quantum error-correcting code will correct
the phase part of the errors.  If there are any bit flip errors left
over, they will almost certainly be detected in the inner layer of
encoding by the classical authentication scheme.  Since it is just a
classical authentication scheme, however, residual phase errors will
{\em not} be detected.  Therefore, we want our quantum
error-correcting code to have the property that for any $E$, for most
$E'$s produced by Hadamard transforms, either the code will correct
both bit and phase flip errors, or it will leave uncorrected bit flip
errors.  That is, it should (almost) never correct all bit flip errors
but leave uncorrected phase errors.

One way (though possibly not the only way) to accomplish this is to
let the phase error correction rate be slightly higher than the bit
flip error correction rate.  That is, we choose a CSS code that
corrects a fraction $\delta$ of bit flip errors and a fraction
$\delta+\eta$ of phase errors.  In section~\ref{sec:prepandmeas}, we
chose two classical codes $C_1$ and $C_2$.  The CSS code which uses
$C_1$ to correct bit flips and $C_2$ to correct phase flips will have
the requisite properties.

We can take this purity testing code and produce a quantum
authentication scheme as per Barnum {\it et al}~\cite{qas}.  To do
this, we take the quantum state, encrypt it, then further encode
it using the above purity testing code with random syndromes.
Encrypting the quantum state means performing a random Pauli
matrix (determined by part of the shared secret key) on each
qubit. However, when we move to an uncloneable encryption scheme,
we are simply sending classical basis states.  Therefore, an
initial phase randomization step has no effect on the state.  We
thus get the following uncloneable encryption protocol for an
$n$-bit message, based on the classical key $(k, e, c_1, c_2, b)$:
\begin{enumerate}
\item Divide the $n$ input bits into $r$ groups of size $s$.  View
each group of $s$ bits as an element of the finite field $GF(2^s)$.
$k$ is a string of size $s$, which we can also view as an element of
$GF(2^s)$.
\item The $r$ resulting registers $m_0, \ldots, m_{r-1}$ can be viewed
as the first $r$ coefficients of a degree $r$ polynomial $f(z)$.  The
final coefficient $m_r$, the constant term, will be chosen so that
$f(k) = 0$.  That is, $\sum_i m_i k^{r-i} = 0$.
\item Alice XORs the string $(m_0, \ldots, m_r)$ with the $(n+s)$-bit
string $e$, producing a new classical string $y$ of length $n+s$ bits.
\item Alice creates the basis state $\ket{y}$ and encodes it using the
CSS code $\{C_1, C_2\}$ with syndrome $c_1$ for $C_1$ and $c_2$ for
$C_2$.
\item Alice takes the resulting $N$-qubit state and performs a
Hadamard transform on it for each location where the $N$-bit string
$b$ is a $1$.  The resulting state $\ket{\psi}$ Alice transmits to
Bob.
\end{enumerate}

To decode, Bob simply takes the state he receives, reverses the
Hadamard transforms, corrects and decodes the state from the quantum
error-correcting code, and measures the resulting state (which should
be a basis state if all has gone correctly).  He then XORs the
resulting classical string $y$ with $e$ and divides the result into
$r+1$ $s$-bit registers, viewed as elements of $GF(2^s)$ --- in fact,
as coefficients of a degree $r$ polynomial over $GF(2^s)$.  He then
evaluates this polynomial at the point $k \in GF(2^s)$, and accepts
the message only if he gets $0$ as the result.  If he does accept, the
message is the first $n$ bits of $y$.

By theorem~\ref{thm:qas}, this is a perfectly good uncloneable
encryption scheme.  Unfortunately, the use of quantum
error-correcting codes means it is not a ``prepare-and-measure''
scheme.  However, following Shor and Preskill~\cite{SP}, we can
note that the phase error correction does not influence the final
outcome, so need not be performed.  In fact, Alice could simply
measure the state before sending, and it would not influence Eve's
attack or Bob's decoding at all.  For a more detailed discussion
of the Shor and Preskill technique, see~\cite{NC,twoway,SP,cont}.
The end result of this procedure is to give us the
``prepare-and-measure'' protocol presented in
section~\ref{sec:prepandmeas}.


\begin{thebibliography}{99}
%
\bibitem{schneier} B.~Schneier, {\it Applied Cryptography} (John Wiley
\& Sons, 1996).
%
\bibitem{WZ} W.~K.~Wootters and W.~H.~Zurek, ``A single quantum cannot
be cloned,'' Nature {\bf 299}, 802 (1982).
%
\bibitem{Dieks} D.~Dieks, ``Communication by EPR devices,''
Phys.\ Lett.\ A {\bf 92}, 271 (1982).
%
\bibitem{NC}  M.~A.~Nielsen and I.~L.~Chuang, {\it Quantum Computation and
Quantum Information} (Cambridge University Press, 2000).
%
\bibitem{PT} D. Gottesman and H.-K. Lo, ``From quantum cheating to
quantum security,'' {\it Physics Today} {\bf 53}, No.~11, p.~22
(Nov.~2000), quant-ph/0111100.
%
\bibitem{BB84}  C. H. Bennett and G. Brassard, ``Quantum
cryptography: public key distribution and coin tossing,''  in
{\it Proceedings of IEEE International Conference on Computers,
Systems, and Signal Processing}, p.~175 (IEEE press, 1984).
%
\bibitem{sdc} A.~Beige, B.-G.~Englert, C.~Kurtsiefer, H.~Weinfurter,
``Secure communication with a publicly known key,'' Acta Phys.\ Pol.\
A {\bf 101}, 357 (2002), quant-ph/0111106.
%
\bibitem{BBB} C.~H.~Bennett, G.~Brassard, S.~Breidbart,
``Quantum Cryptography II: How to re-use a one-time pad safely
even if P=NP,''  unpublished manuscript, November 1982.
%
\bibitem{qenc1} A.~Ambainis, M.~Mosca, A.~Tapp and R.~de Wolf,
``Private quantum channels'', in {\em Proceedings of the 41st Annual
Symposium on Foundations of Computer Science}, p.~547 (IEEE Computer
Society Press, 2000), quant-ph/0003101.
%
\bibitem{qenc2} P.~O.~Boykin, V.~Roychowdhury, ``Optimal encryption of
quantum bits,'' quant-ph/0003059.
%
\bibitem{qas} H.~Barnum, C.~Crepeau, D.~Gottesman, A.~Smith, and
A.~Tapp, ``Authentication of quantum messages,'' quant-ph/0205128, to
appear in Proc.\ FOCS '02.
%
\bibitem{mayers} D. Mayers, ``Unconditional security in quantum
cryptography,'' Journal of ACM {\bf 48}, 351 (2001), quant-ph/9802025.
%
\bibitem{twoway} D.~Gottesman and H.-K.~Lo, ``Proof of security of
quantum key distribution with two-way classical communications,''
quant-ph/0105121, to appear in IEEE Trans.\ Info.\ Theory.
%
\bibitem{B92} C.\,H.~Bennett, ``Quantum cryptography using any two
nonorthogonal states,'' Phys.\ Rev.\ Lett. {\bf 68}, 3121 (1992).
%
\bibitem{SP}  P. W. Shor and J. Preskill, ``Simple proof of
security of the BB84 quantum key distribution protocol,''
Phys.\ Rev.\ Lett.\ {\bf 85}, 441 (2000), quant-ph/0003004.
%
\bibitem{CS} A. R.~Calderbank and P. W.~Shor, ``Good quantum
error-correcting codes exist'', Phys. Rev. A {\bf 54}, 1098 (1996),
quant-ph/9512032.
%
\bibitem{cont} D.~Gottesman and J.~Preskill, ``Secure quantum key
distribution using squeezed states,'' Phys.\ Rev.\ A {\bf 63}, 022309,
18 pages (2001), quant-ph/0008046.
%
\bibitem{maurer} U.~Maurer, ``Conditionally-perfect secrecy and a
provably-secure randomized cipher,'' Journal of Cryptology {\bf 5},
53 (1992).
%
\bibitem{rabin1} Y.~Aumann and M.~O.~Rabin, ``Information
theoretically secure communication in the limited storage space
model,'' in {\it Advances in Cryptology --- CRYPTO '99}, Lecture Notes
in Computer Science {\bf 1807}, p.~65 (Springer-Verlag, Heidelberg,
1999).
%
\bibitem{rabin2} Y.~Z.~Ding and M.~O.~Rabin, ``Hyper-encryption and
everlasting security,'' in STACS 2002, eds.\ H.~Alt and A.~Ferreira,
Lecture Notes in Computer Science {\bf 2285}, p.~1 (Springer-Verlag,
Heidelberg, 2002).
%
\bibitem{efficient} H.-K.~Lo, H.~F.~Chau, and M.~Ardehali, ``Efficient
quantum key distribution scheme and proof of its unconditional
security,'' quant-ph/0011056.
%
\bibitem{HOreuse} J.~Oppenheim, M.~Horodecki, ``How to reuse a
one-time pad and other notes on authentication, encryption and
protection of quantum information,'' quant-ph/0306161.
%
\bibitem{HLMreuse} P.~Hayden, D.~Leung, D.~Mayers, in preparation
(2003).
%
\bibitem{stabintro} D.~Gottesman, ``An introduction to quantum error
correction,'' in {\it Quantum Computation: A Grand Mathematical
Challenge for the Twenty-First Century and the Millenium}, ed.\
S.~J.~Lomonaco, Jr., p.~221 (American Mathematical Society,
Providence, 2002), quant-ph/0004072.
%
\bibitem{Steane} A.~Steane, ``Multiple particle interference and
quantum error correction,'' Proc.\ Roy.\ Soc.\ Lond.\ A{\bf 452}, 2551
(1996), quant-ph/9601029.
%
\end{thebibliography}
\end{document}